\documentclass{ifacconf}

\usepackage{amsmath,amssymb,amsfonts}
\usepackage{algorithmic}
\usepackage{graphicx}
\usepackage{algorithm}
\usepackage{textcomp}
\usepackage{natbib}        

\usepackage{bm, mathtools, marvosym, accents}
\usepackage{epstopdf}
\usepackage{tikz, environ, float, xcolor}
\usetikzlibrary{shapes, arrows, calc, patterns, decorations.pathreplacing, backgrounds, positioning, fit}
\usepackage{enumitem} 
\usepackage{stfloats}

\usepackage{setspace}

\newtheorem{theorem}{Theorem}

\newtheorem{assumption}{Assumption}

\newtheorem{example}{Example}
\newtheorem{remark}{Remark}
\newtheorem{proposition}{Proposition}

\newcommand{\argmin}{\operatornamewithlimits{argmin}}

\begin{document}
\begin{frontmatter}

\title{Feasible-Set Reshaping for Constraint Qualification in Optimization-Based Control\thanksref{footnoteinfo}} 

\thanks[footnoteinfo]{This work was supported in part by in part by NSFC grants 62325303, 62333004 and 62521001, and in part by NSF grant ECCS-2210320.}

\author[First]{Si Wu} 
\author[First]{Tengfei Liu} 
\author[Second]{Yiguang Hong}
\author[Third]{Zhong-Ping Jiang}
\author[First]{Tianyou Chai} 

\address[First]{State Key Laboratory of Synthetical Automation for Process Industries, Northeastern University, Shenyang, 110004, China {\tt\small(e-mails: wusixstx@163.com; tfliu@mail.neu.edu.cn; tychai@mail.neu.edu.cn)}.}
\address[Second]{Department of Control Science and Engineering, Tongji University, Shanghai, China. {\tt\small(e-mail: 	
yghong@tongji.edu.cn)}}
\address[Third]{Department of Electrical and Computer Engineering, New York University, Brooklyn, NY 11201, USA {\tt\small (e-mail: zjiang@nyu.edu)}}

\begin{abstract}
This paper presents a novel feasible-set reshaping technique to optimization-based control with ensured constraint qualification. In our problem setting, the feasible set of admissible control inputs depends on the real-time state of the plant, and the linear independence constraint qualification (LICQ) may not be satisfied in some regions of interest. By feasible-set reshaping, we project the constraints of the original feasible set onto an appropriately chosen constant matrix with its rows forming a positive span of the space of the optimization variable. It is proved that the reshaped feasible set is nonempty and satisfies LICQ, as long as the original feasible set is nonempty. The effectiveness of the proposed method is verified by constructing Lipschitz continuous quadratic-program-based (QP-based) controllers based on the reshaped feasible sets.
\end{abstract}


\end{frontmatter}

\section{Introduction}

Optimization-based control is promising for systems subject to constraints on the state and control input. Such controllers first determine the set of admissible control inputs consistent with the constraints, and then select from this set a control input that drives the system toward primary control objectives while ensuring constraint satisfaction \citep{Escande-Mansard-Wieber-IJRR-2014,Wang-Ames-Egerstedt-TRO-2017,Angeliki-Ioannis-Antonios-Ioannis-2020-SS}. Methods based on control barrier functions (CBFs) and control Lyapunov functions (CLFs) have proven effective in addressing nonlinear constraints on the state of a control-affine system by imposing constraints affine in control on the control inputs, thereby enabling the application of nonlinear control techniques to constrained control problems. As a representative case, safety-critical control has recently attracted significant attention \citep{Xu-Tabuada-Grizzle-Ames-IFAC-2015,Ames-Xu-Grizzle-Tabuada-TAC-2017}.

When the plant is subject to uncertain dynamics and state-dependent constraints, robustness becomes a central concern \citep{Jankovic-Auto-2018,Mestres-Cortes-Auto-24}. Lipschitz continuity of the closed-loop dynamics is crucial for ensured robustness of a control system \citep{Teel-Praly-ESAIM-00}. Therefore, in the context of optimization-based control, it often requires that the solution to the underlying optimization problem varies smoothly with respect to the constraints, specifically exhibiting at least a locally Lipschitz continuous manner, for robustness analysis.

The linear independence constraint qualification (LICQ) \citep{Peterson-SIAMReview-1973}, which means that the gradients of the active constraints with respect to the optimization variable are linearly independent, is commonly used to guarantee Lipschitz continuity of solutions to parametric optimization problems arising in optimization-based controllers. \citet{Fiacco-1976-MP} explicitly constructed the partial derivatives of local solutions to parametric nonlinear programs under LICQ and other mild regularity conditions. \citet{Hager-SIAMControl-1979} establishes the Lipschitz continuity of parametric convex programs, under the assumption that the gradients of the binding constraints are linearly independent, which can be guaranteed by LICQ. The Lipschitz continuity results for parametric optimization have subsequently been used in the continuity analysis of optimization-based controllers \citep{Morris-Powell-Ames-CDC-2015}.
However, within the context of optimization-based control, LICQ is generally difficult to guarantee a priori, as the set of admissible control inputs typically depends on and varies with the state of the plant. This limitation has been noted by \citet{Robinson-1982}, revisited in \citet{Mestres-Allibhoy-Cortes-EJC-2025}, and illustrated through safety-oriented counterexamples in \citet{Wu-Liu-Egerstedt-Jiang-2023-TAC}. To date, there seems no systematic method to ensure LICQ for optimization-based control.

In our previous studies \citep{Wu-Liu-Egerstedt-Jiang-2023-TAC,Wu-Liu-Niu-Jiang-RAL-2022,Wu-Liu-Zhang-Ding-Jiang-Chai-TAC-25}, we developed feasible-set reshaping techniques that employ carefully selected constant constraint matrices to represent half-space (affine) constraints in a way that guarantees the satisfaction of the LICQ condition. \citet{Agrawal-Lee-Panagou-ARXIV-25} reconstructed a single second-order cone constraint from multiple state-dependent half-space constraints, assuming that the original constraint matrix has unit-norm rows. In our most recent work \citep{Wu-Liu-Jiang-2025-TAC}, we further extended the feasible-set reshaping approach to handle constraints involving state-independent quadratic terms. To date, these advancements remain largely restricted to half-space constraints or certain classes of quadratic constraints. This motivates the investigation presented in this paper.

This paper presents a nontrivial refinement of the feasible-set shaping method. Specifically, given a closed convex feasible set defined by multiple quadratic constraints, our method projects the original constraints onto a carefully chosen constant matrix. The matrix is chosen such that its rows form a positive span of the control input space, and any submatrix with the number of rows no larger than the dimension of the optimization variable is full row rank. It is ensured that the reshaped feasible set is nonempty and satisfies LICQ. The effectiveness of the proposed method is verified by constructing Lipschitz continuous quadratic-program-based (QP-based) controllers using the reshaped feasible sets.

{\bfseries Preliminaries and Notations.}

In this paper, $|\cdot|$ represents the Euclidean norm for real vectors and the induced $2$-norm for real matrices. For a real matrix $A$, we use $[A]_{i,j}$, $[A]_{i,:}$ and $[A]_{:, j}$ to represent the element at row $i$ and column $j$, the $i$-th row vector and the $j$-th column vector, respectively. For a vector $v$, we use $[v]_i$ to represent the $i$-th element of the vector $v$. For real column vectors $x_1,\ldots,x_m$, we use $(x_1;\ldots;x_m)$ to denote $(x_1^T,\ldots,x_m^T)^T$. For a symmetric matrix $A\in\mathbb{R}^{n\times n}$, $\lambda_{\max}(A)$ and $\lambda_{\min}(A)$ denote the maximum eigenvalue and the minimum eigenvalue, respectively. 
Denote $\mathbb{S}^{n-1}=\{x\in\mathbb{R}^n:|x|=1\}$.

Following the standard treatment in \cite{Boyd-Vandenberghe-book-2004}, we briefly recall some concepts in the field of optimization. A half-space in $\mathbb{R}^n$ is a set of the form $\{ u \in \mathbb{R}^{n_u} \mid a^T u \leq b \}$, where $a \neq 0$. A constraint is active at a feasible point if it holds with equality, and inactive if it holds as a strict inequality. 

\section{Problem Formulation} \label{section_problemformulation}

This section introduces a class of nonlinear convex optimization problems that commonly arise in optimization-based controller design and formulates the corresponding constraint qualification problem.

Consider a plant modeled by 
\begin{align}
\dot{x} = f(x,u) \label{eq_plant}
\end{align}
where $x\in\mathbb{R}^{n_x}$ is the state, $u\in\mathbb{R}^{n_u}$ is the control input and $f:\mathbb{R}^{n_x}\times\mathbb{R}^{n_u}\to\mathbb{R}^{n_x}$ is a locally Lipschitz function with $f(0,0)=0$. 

In optimization-based control systems, the feasible set of admissible control inputs is often defined as
\begin{align}
\mathcal{U}(x) &= \{u\in\mathbb{R}^{n_u}:g_i(x,u) \le 0,~ i=1,\ldots,n_c\},  \label{eq_def_original_feasibleset}
\end{align}
where each function $g_i:\mathcal{X}\times \mathbb{R}^{n_u}\to \mathbb{R}$ is given by
\begin{align}
g_i(x,u) &= \frac{1}{2} u^T P_i(x) u + q_i^T(x)u + r_i(x) \label{eq_def_fi}
\end{align}
with $P_i\in\mathcal{X}\to\mathbb{R}^{n_u\times n_u}$ being a locally Lipschitz, positive semidefinite function, $q_i:\mathcal{X}\to\mathbb{R}^{n_u}$ and $r_i:\mathcal{X}\to\mathbb{R}^{n_u}$ being locally Lipschitz functions. The set $\mathcal{X}\subseteq\mathbb{R}^{n_x}$ is assumed to be nonempty and open, representing the region of interest for the state $x$.

The following assumption is made on $\mathcal{U}(x)$ and $g_i(x,u)$.

\begin{assumption} \label{assumption_original_feasibleset}
There exists a function 
$u_s:\mathcal{X}\to\mathbb{R}^{n_u}$ such that
\begin{align}
u_s(x) \in \mathcal{U}(x), \quad \forall x\in\mathcal{X}. \label{eq_assumption_selection}
\end{align}
Moreover, for all $i=1,\ldots,n_c$, it holds that
\begin{align}
\frac{\partial g_i(x,u)}{\partial u} \neq 0, \quad \forall u\in\mathbb{R}^{n_u}, x\in\mathcal{X}.
\end{align}
\end{assumption}

\begin{remark}
In typical scenarios of optimization-based control, the feasible set $\mathcal{U}(x)$ is often constructed as a closed convex set. For instance, CBF-based methods can ensure that control-affine systems satisfy state constraints by restricting the control input to a closed convex set \cite{Jankovic-Auto-2018, Ames-Coogan-Egerstedt-ECC-2019}. In addition, various saturation constraints on the control input can be modeled as a closed convex set \cite{Cortez-Dimarogonas-Auto-2022, Cortez-Tan-Dimarogonas-2021-CSL}. Note that one can always reconstruct a closed convex feasible set by appropriately shrinking the original feasible set, as the closure of any set in a metric space is always a closed set.
\end{remark}

Consider the plant \eqref{eq_plant} and the original feasible set $\mathcal{U}(x)$ in \eqref{eq_def_original_feasibleset}. Under Assumption \ref{assumption_original_feasibleset}, for each $x\in\mathcal{X}$, we aim to construct a reshaped feasible set $\mathcal{U}_L(x)$ based on $\mathcal{U}(x)$ such that:
\begin{itemize}
\item $\mathcal{U}_L(x)$ is nonempty and satisfies $\mathcal{U}_L(x)\subseteq \mathcal{U}(x)$; 
\item LICQ holds at any feasible point $u_* \in \mathcal{U}_L(x)$ for which the number of active constraints is no greater than $n_u$, that is, at $u_*$, the gradients of the active constraints with respect to the optimization variable are linearly independent.
\end{itemize}

In optimization-based control problems, the plant is typically required to achieve primary control objectives (such as reference tracking), while strictly satisfying all imposed constraints. Following the convention in \citep{Ames-Xu-Grizzle-Tabuada-TAC-2017}, we consider the following QP-based controller: 
\begin{align}
u = &\argmin_{u\in\mathcal{U}(x)} |u - u_c|^2, \label{eq_optimization_original}
\end{align}
where $u_c\in\mathbb{R}^{n_u}$ is the primary control input representing the primary control objective, and the control input $u\in\mathbb{R}^{n_u}$ is generated as the optimization variable. 

Note that, for a QP with the reshaped feasible set $\mathcal{U}_L(x)$ and a cost function $|u - u_c|^2$, if LICQ holds at any feasible point $u_* \in \mathcal{U}_L(x)$ for which the number of active constraints is no greater than $n_u$, it can be shown that the solution to the QP is Lipschitz continuous with respect to the data that define the optimization, using the existing approach to parametric convex programs \citep{Hager-SIAMControl-1979}.

The following example illustrates a case where the LICQ condition is violated, which results in a non-Lipschitz continuous optimization-based controller.

\begin{example}\label{example_01}
Consider the plant $\dot{x}=u$ with $x\in\mathbb{R}^2$ and $u\in\mathbb{R}^2$, and the optimization-based controller \eqref{eq_optimization_original} defined with the primary control input  
\begin{align}
u_c(t)=[1;0],\qquad t\ge 0 \label{eq_simulation1_uc}
\end{align}  
and the feasible set $\mathcal{U}(x)$ in the form of \eqref{eq_def_original_feasibleset} with 
\begin{subequations} \label{eq_simulation1_Pqr}
\begin{align}
P_i(x) &= 0_{2\times 2}, \qquad r_i(x)=0, \quad \forall i=1,2,\\
q_1(x) &= \frac{1}{\sqrt{1+\sin^2 [x]_1}}
\begin{bmatrix}
-\sin x_1\\
1
\end{bmatrix}, \\
q_2(x) &= \frac{1}{\sqrt{1+\sin^2 [x]_1}}
\begin{bmatrix}
-\sin x_1\\
-1
\end{bmatrix}
\end{align}
\end{subequations}  
for $x\in\mathbb{R}^2$. When $[x]_1 = k\pi$ with $k \in \mathbb{Z}$, the feasible set $\mathcal{U}(x)$ can be rewritten as
\begin{align}
\mathcal{U}(x) = \{u \in \mathbb{R}^2 : [0; 1]^Tu \leq 0, [0;-1]^Tu \leq 0\}. \label{eq_example_Ux}
\end{align}
Since the vectors $[0;1]^T$ and $[0;-1]^T$ are linearly dependent, LICQ fails to hold at any feasible point in $\mathcal{U}(x)$ defined in \eqref{eq_example_Ux}. Consequently, the optimization-based controller may not be Lipschitz continuous with respect to the state $x$.

Figure \ref{simulation_u_norm01} shows $|u|$ obtained by solving the QP \eqref{eq_optimization_original} with the original feasible set $\mathcal{U}(x)$ for $[x]_1\in[-2.5\pi,1.5\pi]$ and $[x]_2\in[-1,1]$. It can be observed that $|u|$ exhibits sudden changes when $[x]_1=k\pi$ for $k\in\mathbb{Z}$.

\begin{figure}[!ht]
\centering
\includegraphics[width=0.8\linewidth]{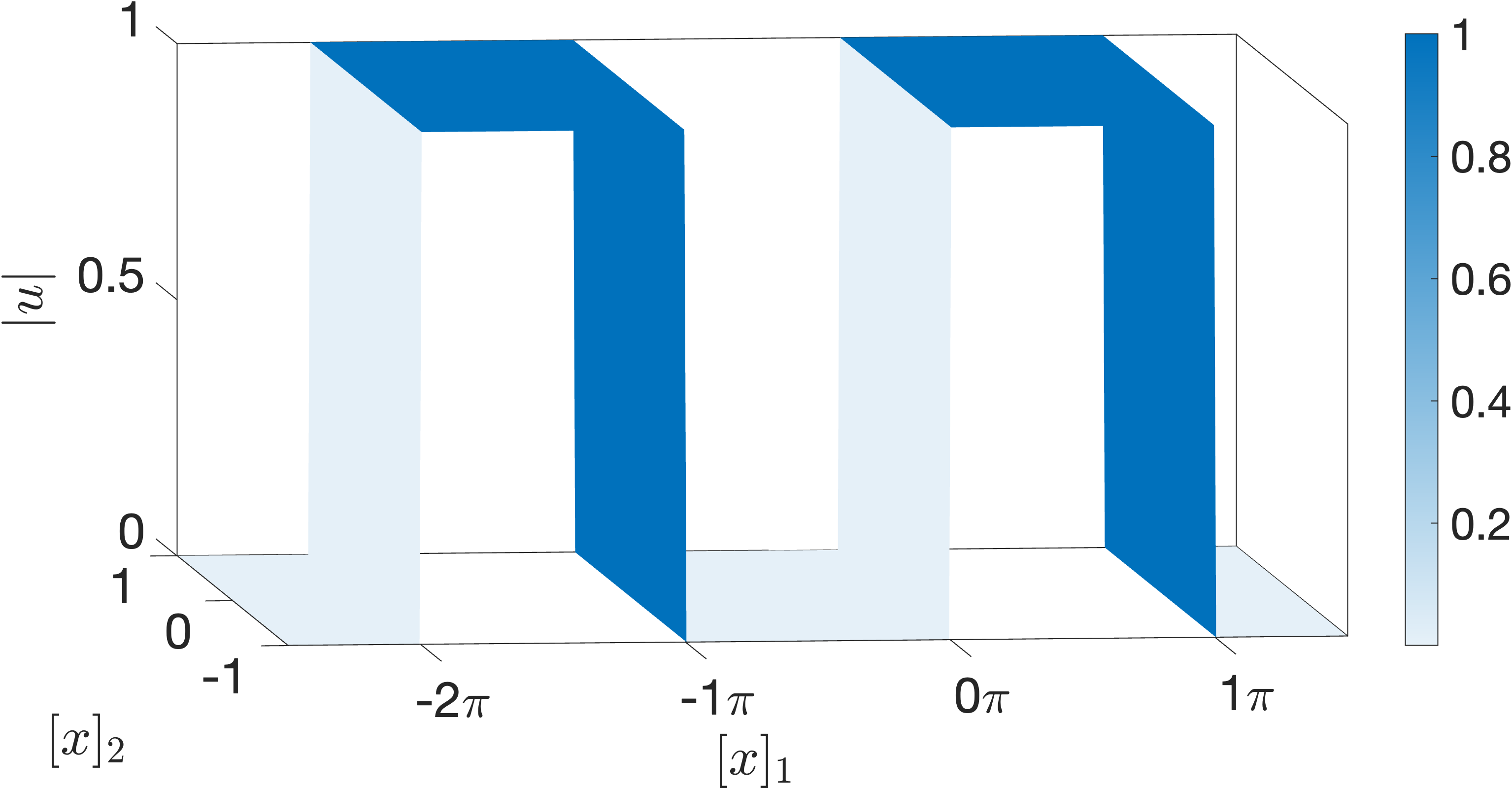}
\caption{The solution $u$ of a QP exhibits sudden changes with respect to $x$.}
\label{simulation_u_norm01}
\end{figure}
\end{example}

\section{The Feasible-Set Reshaping Technique} \label{section_feasibleset_reshaping}

In this section, we propose a method to reshape the original feasible set defined by multiple quadratic constraints, such that the resulting reshaped feasible set is a subset of the original one and satisfies the LICQ condition at every feasible point. Furthermore, leveraging this feasible-set reshaping technique, we design a QP-based controller that admits boundedness and Lipschitz continuity properties.

\subsection{Half-Space Representation of the Original Feasible Set} \label{subsetction_representation}

We begin by equivalently representing the original feasible region as a collection of half-space constraints.

Consider the original feasible set $\mathcal{U}(x)$ defined in \eqref{eq_def_original_feasibleset}, where the functions $g_i:\mathcal{X}\times \mathbb{R}^{n_u}\to\mathbb{R}$ for $i=1,\ldots,n_c$ are given in \eqref{eq_def_fi}. The following proposition shows that $\mathcal{U}(x)=\mathcal{U}_a(x)$, where 
\begin{subequations} \label{eq_linearized_feasible_set_all}
\begin{align}
\mathcal{U}_a(x) &= \bigcap_{i=1,\ldots,n_c}\bigcap_{\nu\in\mathbb{R}^{n_u}} \mathcal{H}_{i,\nu}(x), \label{eq_linearized_feasible_set}\\
\mathcal{H}_{i,\nu}(x) &= \left\{u\in\mathbb{R}^{n_u} : \frac{\partial g_i(x,\nu)}{\partial \nu} (u-\nu)\le -g_i(x,\nu)\right\}. \label{eq_halfspace}
\end{align}
\end{subequations}
The following proposition also gives a normalized representation of $\mathcal{H}_{i,\nu}(x)$, and explicitly constructs a feasible point $u_s(x)$ of $\mathcal{U}(x)$.

\begin{proposition}\label{prop_properties_Ua}
Given the original feasible set $\mathcal{U}(x)$ defined in \eqref{eq_def_original_feasibleset} with functions $g_i(x,u)$, $P_i(x)$, $q_i(x)$ and $r_i(x)$ for $i=1,\ldots,n_c$, consider the set $\mathcal{U}_a(x)$ defined by \eqref{eq_linearized_feasible_set_all}. For any $x\in\mathcal{X}$, the following properties hold:
\begin{itemize}
\item \textbf{Equivalent representation:} $\mathcal{U}(x) = \mathcal{U}_a(x)$.
\item \textbf{Normalized representation of $\mathcal{H}_{i,\nu}(x)$:} Under Assumption \ref{assumption_original_feasibleset}, $\mathcal{H}_{i,\nu}(x)$ defined by \eqref{eq_halfspace} can be equivalently represented in a normalized form:
\begin{subequations} \label{eq_proposition_normalized_representation} 
\begin{align}
\mathcal{H}_{i,\nu}(x) &= \{u\in\mathbb{R}^{n_u} : A_{i,\nu}(x)(u-u_s(x)) \notag \\
&~~~~~~~~~~~~~~~~~~\le b_{i,\nu}(x)\}, \label{eq_halfspace_new} \\
A_{i,\nu}(x) &= \frac{\partial g_i(x,\nu)}{\partial \nu}\left|\frac{\partial g_i(x,\nu)}{\partial \nu}\right|^{-1}, \\
b_{i,\nu}(x) &= A_{i,\nu}(x)(\nu-u_s(x))\notag \\
&~~~-g_i(x,\nu)\left|\frac{\partial g_i(x,\nu)}{\partial \nu}\right|^{-1}.
\end{align}
\end{subequations}

\item \textbf{A feasible point of $\mathcal{U}(x)$:} If $P_i(x)=0$, $q_i^T(x)\neq 0$ and
\begin{align} 
\frac{r_i(x)}{|q_i(x)|}+\frac{r_j(x)}{|q_j(x)|} \le 0 \label{eq_proposition_feasible_point_condition}
\end{align}
holds for all $i,j=1,\ldots,n_c$, $i\neq j$, then one can construct an element $u_s(x) \in \mathcal{U}(x)$ as 
\begin{align}
u_s(x) = \sum^{n_c}_{i=1} \frac{q_i(x)}{|q_i(x)|^2}\min\{-r_i(x), 0\}. \label{eq_proposition_feasible_point}
\end{align}
\end{itemize}
\end{proposition}

The proof of Proposition \ref{prop_properties_Ua} is postponed to Appendix \ref{proof_prop_properties_Ua}.

\subsection{Projecting Each Half-Space Constraint onto the Rows of an Appropriately Chosen Matrix $A_L$} \label{subsection_projecting}

We construct a function to reshape each half-space constraint $\mathcal{H}_{i,\nu}$ defined in \eqref{eq_proposition_normalized_representation} based on the predetermined constant matrix $A_L$, such that the resulting set belongs to $\mathcal{H}_{i,\nu}$. 

Given a constant $c_A$ with $0 < c_A < 1$,  $A_L$ is chosen to satisfy the following conditions: 
\begin{enumerate}[label=(\alph*), ref=(\alph*)]
\item \label{positive_basis_normalization} For $i=1,\ldots,n_l$, $[A_L]_{i,:}$ is a unit vector;
\item \label{positive_basis_independent} Any $n_u$ rows of $A_L$ are linearly independent; 
\item \label{positive_basis_combination} Any unit vector $l_0\in\mathbb{R}^{n_{x_2}}$ can be represented as a positive combination\footnote{A positive combination of vectors $\{l_i\in\mathbb{R}^n: i=1,\ldots,m\}$ is a linear combination $a_1l_1 + \cdots + a_ml_m$ with $a_i\ge 0$ for $i=1,\ldots,m$; it is called strictly positive if $a_i > 0$ for all $i$. See \cite{Davis-AJM-1954} for details.} of the vectors in the set $\mathcal{L}(l_0)$, where 
\begin{align}
\mathcal{L}(l_0)=\{[A_L]_{i,:}^T : [A_L]_{i,:}l_0 \ge c_A, i = 1,\ldots, n_l\}, \label{eq_set_Ll0}
\end{align}
represents the set of transposed rows of $A_L$ whose inner product with $l_0$ is no less than $c_A$, with cardinality no less than $n_u$. 
\end{enumerate}
Given $0<c_A<1$, \cite{Wu-Liu-Niu-Jiang-RAL-2022, Wu-Liu-Egerstedt-Jiang-2023-TAC} already showed the existence of a matrix $A_L$ satisfying the above conditions.

Consider each half-space constraint $\mathcal{H}_{i,\nu}$ defined in \eqref{eq_proposition_normalized_representation} for $i=1,\ldots,n_c$ and $\nu\in\mathbb{R}^{n_u}$. 
We project the vector $A^T_{i,\nu}(x)b_{i,\nu}(x)$ associated with the half-space $\mathcal{H}_{i,\nu}(x)$ onto the directions of the adjacent row vectors in $A_L$, and then activate the corresponding components along the remaining row directions. In particular, we introduce the following function to combine the projection and the activation:
\begin{align}
\phi(b,A^T)
&=\underbrace{c_Ab+b\varphi(A_LA^T-c_A)}_{\text{Projection}}
+\underbrace{k_{\phi}\varphi(c_A-A_LA^T)}_{\text{Activation}} \label{eq_activation_functions}
\end{align}
for $b\in\mathbb{R}_+$ and $A\in \mathbb{S}^{n_u-1}$, where $k_{\phi}$ is a positive constant, and $\varphi:\mathbb{R}^{n_l}\to\mathbb{R}_+^{n_l}$ is defined by $[\varphi(s)]_i=\max\{[s]_i,0\}$ for $i=1,\ldots,n_l$. 

The following proposition gives the explicit form and some important properties of the reshaped feasible set of $\mathcal{H}_{i,\nu}(x)$.

\begin{proposition}\label{proposition_halfspace}
Given a set $\mathcal{H}_{i,\nu}(x)$ defined by $A_{i,\nu}:\mathcal{X}\to\mathbb{S}^{n_u-1}$, $b_{i,\nu}:\mathcal{X}\to\mathbb{R}_+$ and $u_s:\mathcal{X}\to\mathbb{R}^{n_u}$ in \eqref{eq_proposition_normalized_representation} for $i=1,\ldots,n_c$, $\nu\in\mathbb{R}^{n_u}$ and $x\in\mathcal{X}$, the reshaped feasible set of $\mathcal{H}_{i,\nu}(x)$ defined by
\begin{align}
\mathcal{U}_{b'}(x) = \{&u\in\mathbb{R}^{u_b} : A_L(u-u_s(x))\le b'\} \label{eq_reshaped_halfspace}
\end{align}
is nonempty and contains $u_s(x)$ for $x\in\mathcal{X}$ and $b'\in\mathbb{R}^{n_l}_+$. Moreover, for any $b'\in\mathbb{R}^{n_l}_+$ and any $x\in\mathcal{X}$, if 
\begin{align}
b' \le \phi(b_{i,\nu}(x), A^T_{i,\nu}(x)), \label{eq_reshaped_halfspace_condition}
\end{align}
then $\mathcal{U}_{b'}(x) \subseteq \mathcal{H}_{i,\nu}(x)$.
\end{proposition}

The proof of Proposition \ref{proposition_halfspace} is postponed to Appendix \ref{appendix_proof_proposition_halfspace}.

\subsection{Incorporate all the Projected Constraints}

Leveraging the results from Sections \ref{subsetction_representation} and \ref{subsection_projecting}, the reshaped feasible set of $\mathcal{U}(x)$ is constructed as:
\begin{align}
\mathcal{U}_L(x) = \{u\in\mathbb{R}^{n_u}:A_L(u-u_s(x))\le b_L(x)\} \label{eq_reshaped_feasibleset}
\end{align}
where $b_L:\mathcal{X}\to\mathbb{R}^{n_l}_{+}$ is defined by 
\begin{align}
[b_L(x)]_i = \min_{j=1,\ldots,n_c} \inf_{\nu\in\mathbb{R}^{n_u}} [\phi(b_{j,\nu}(x), A_{j,\nu}^T(x))]_i. \label{eq_b_L}
\end{align}

The following theorem shows that the reshaped feasible set is nonempty, belongs to the original feasible set, and admits the LICQ property.

\begin{theorem} \label{theorem}
Consider the plant \eqref{eq_plant} and the original feasible set $\mathcal{U}(x)$ in \eqref{eq_def_original_feasibleset}. Supposed that Assumption \ref{assumption_original_feasibleset} holds with $u_s:\mathcal{X}\to\mathbb{R}^{n_u}$. For each $x\in\mathcal{X}$, the reshaped feasible set $\mathcal{U}_L(x)$ defined in \eqref{eq_reshaped_feasibleset} admits the following properties:
\begin{itemize}
\item \textbf{Feasibility and constraint satisfaction:} The reshaped feasible set $\mathcal{U}_L(x)$ satisfies 
\begin{align}
u_s(x)\in\mathcal{U}_L(x) \subseteq \mathcal{U}(x);
\end{align}
\item \textbf{LICQ:} LICQ holds at any feasible point $u_* \in \mathcal{U}_L(x)$ with the number of active constraints at $u_*$ no larger than $n_u$.
\end{itemize}
\end{theorem}

The proof of Theorem \ref{theorem} is postponed to Appendix \ref{appendix_proof_theorem}. 

Based on the achievement above, we consider constructing an optimization-based controller with the reshaped feasible set $\mathcal{U}_L(x)$ defined in \eqref{eq_reshaped_feasibleset} as 
\begin{align}
u = &\argmin_{\nu\in\mathcal{U}_L(x)} |\nu - u_c|^2. \label{eq_optimization_reshaped}
\end{align} 
Theorem \ref{theorem} guarantees that the QP \eqref{eq_optimization_reshaped} is feasible for all $x\in\mathcal{X}$ and $u_c\in\mathbb{R}^{n_u}$. Denote $\rho_L$ as the map from $(x,u_c)$ to $u$ generated by solving the QP \eqref{eq_optimization_reshaped}. 

The following theorem gives the boundedness and Lipschitz continuity of the controller \eqref{eq_optimization_reshaped}.

\begin{theorem} \label{theorem2}
Consider the plant \eqref{eq_plant}, the original feasible set $\mathcal{U}(x)$ in \eqref{eq_def_original_feasibleset} satisfying Assumption \ref{assumption_original_feasibleset} with $u_s:\mathcal{X}\to\mathbb{R}^{n_u}$, and the optimization-based controller defined in \eqref{eq_optimization_reshaped} with the reshaped feasible set $\mathcal{U}_L(x)$ defined in \eqref{eq_reshaped_feasibleset}. The following properties hold:
\begin{itemize}
\item \textbf{An upper bound:} The function $\rho_L$ satisfies
\begin{align}
|\rho_L(x,u_c)| \le |u_s(x)|+|u_c| \label{eq_theorem_boundedness}
\end{align}
for all $x\in\mathcal{X}$ and $u_c\in\mathbb{R}^{n_u}$;
\item \textbf{Lipschitz continuity:} There exists a positive constant $L>0$ such that 
\begin{align}
&|\rho_L(x_1,u_{c1})-\rho_L(x_2,u_{c2})|\le L( |u_s(x_1)-u_s(x_2)|  \notag \\
&\quad + |u_{c1}-u_{c2}|+|b_L(x_1)-b_L(x_2)|)  \label{eq_theorem_Lipschitz}
\end{align}
for all $x_1,x_2\in\mathcal{X}$ and $u_{c1},u_{c2}\in\mathbb{R}^{n_u}$.
\end{itemize}
\end{theorem}

The proof of Theorem \ref{theorem2} is postponed to Appendix \ref{appendix_proof_theorem2}. 

The following example shows that LICQ holds at any feasible point $u_* \in \mathcal{U}_L(x)$ with the number of active constraints at $u_*$ no larger than $n_u$, and that the optimization-based controller constructed based on the reshaped feasible set is Lipschitz continuous.

\begin{example}\label{example_02}
Continuing Example \ref{example_01}, we apply the proposed feasible-set reshaping technique to the original feasible set. The resulting controller is in the form of \eqref{eq_optimization_reshaped}, where $u_c$ is given in \eqref{eq_simulation1_uc}, $\mathcal{U}_L$ is given in \eqref{eq_reshaped_feasibleset} with
\begin{subequations} \label{eq_example2_ALbL}
\begin{align}
[A_L]_{i,:} &= \left[\cos\frac{2\pi i}{n_l}; \sin\frac{2\pi i}{n_l}\right]^T, \label{eq_example2_AL}\\
[b_L(x)]_{i} &= \min_{j=1,\ldots,2} [\phi(-r_j(x), q_j(x))]_i, \label{eq_example2_bL}\\
u_s(x)  &= \sum^2_{j=1}\frac{q_j(x)}{|q_j(x)|^2}\min\{-r_j(x), 0\}, \\
\phi(b,A^T) &=c_Ab+b\varphi(A_LA^T-c_A) \notag \\
&~~~~~~~~~ +k_{\phi}\varphi(c_A-A_LA^T), \\
k_{\phi} &= 1.2, \quad c_A = \frac{2\pi}{n_l}, \quad n_l=7
\end{align} 
\end{subequations}
for $i=1,\ldots,n_l$, where $q_j(x)$ and $r_j(x)$ for $j=1,2$ are defined in \eqref{eq_simulation1_Pqr}. 

It can be verified that any two rows of $A_L$ defined in \eqref{eq_example2_AL} are linearly independent. Hence, LICQ holds at any feasible point $u_* \in \mathcal{U}_L(x)$ with the number of active constraints at $u_*$ no larger than $2$. This verifies the effectiveness of the feasible-set reshaping technique.

\begin{figure}[!ht]
\centering
\includegraphics[width=0.8\linewidth]{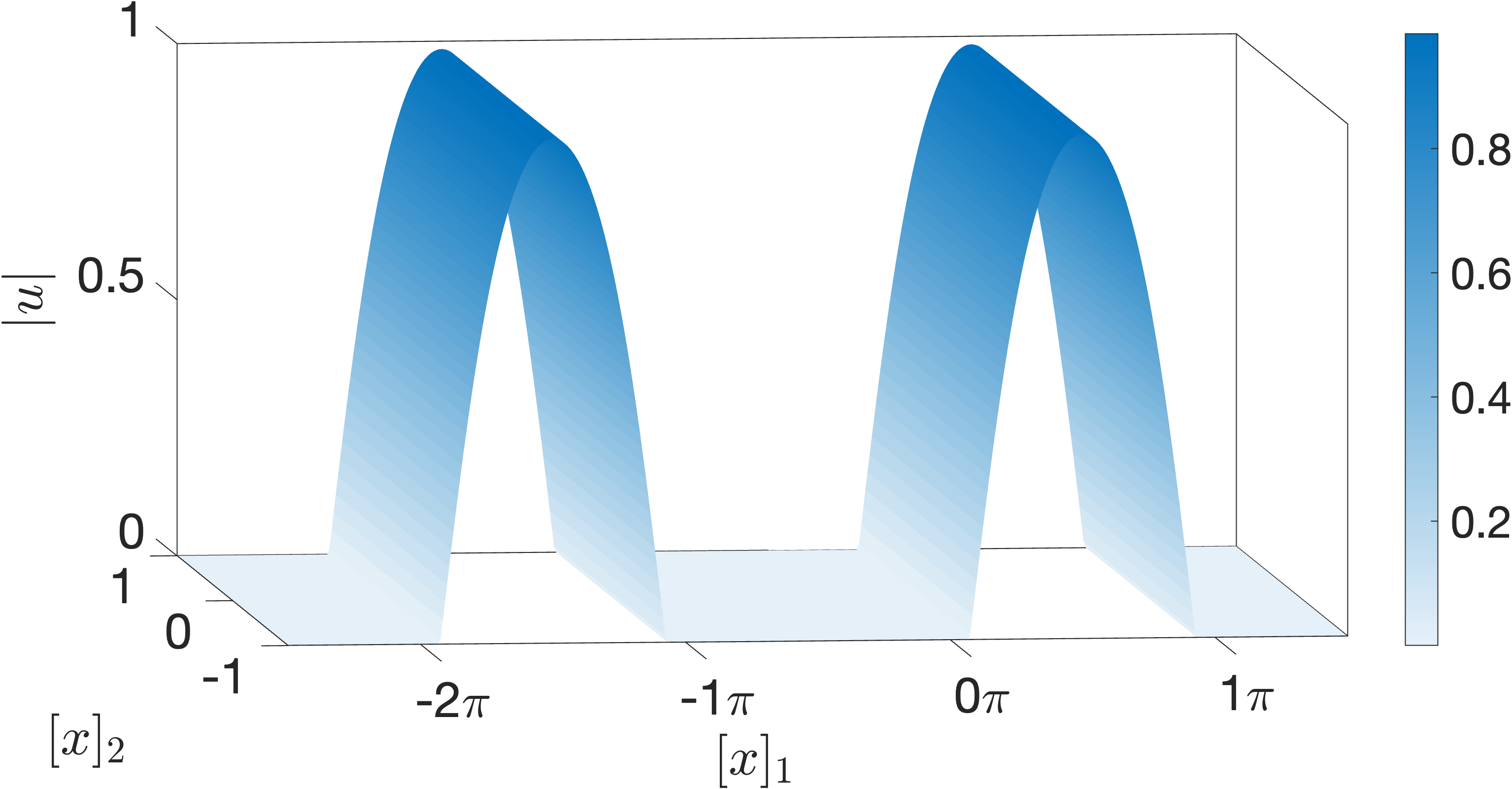}
\caption{The solution of a QP with a reshaped feasible set does not exhibit sudden changes.}
\label{simulation_unorm_Lipschitz}
\end{figure}
\end{example}

Figure \ref{simulation_unorm_Lipschitz} shows $|u|$ obtained by solving the QP \eqref{eq_optimization_reshaped} with the reshaped feasible set $\mathcal{U}_L$ for $[x]_1\in[-2.5\pi,1.5\pi]$ and $[x]_2\in[-1,1]$. Compared with the result in Example \ref{example_01}, the value of $|u|$ changes continuously with respect to the state. This verifies the Lipschitz continuity of the proposed optimization-based controller with the reshaped feasible set.

\begin{figure*}[!ht]
\centering
\includegraphics[width=\textwidth]{materials/main_ifac03_safety_steps2.png}
\caption{Comparison of the original feasible set $\mathcal{U}(x)$ and the reshaped feasible set $\mathcal{U}_L(x)$, together with the corresponding values of $\rho(x,u_c)$ and $\rho_L(x,u_c)$ for selected values of $x$.}
\label{main_ifac03_safety_steps2}
\end{figure*}

\section{Numerical Simulation} \label{section_simulation}

We consider a safety-critical control scenario to verify the effectiveness of the proposed method in addressing the LICQ and Lipschitz continuity issues with optimization-based controllers. In particular, we consider a mobile robot modeled by an integrator in $\mathbb{R}^2$ and two stationary disc-shaped obstacles with centers located at $o_1=[0;-2]$ and $o_2=[0;2]$. 

Define the desired safe set
\begin{align}
\mathcal{S}=\{x\in\mathbb{R}^2:|x-o_i|\ge 2,\ i=1,2\}.
\end{align}
Given the primary control input defined in \eqref{eq_simulation1_uc}, motivated by \cite{Ames-Coogan-Egerstedt-ECC-2019}, we first consider a safety-critical controller in the form of \eqref{eq_optimization_original} with the primary contol input $u_c$ defined in \eqref{eq_simulation1_uc} and the original feasible set $\mathcal{U}(x)$ defined in \eqref{eq_def_original_feasibleset} with
\begin{subequations} \label{eq_simulation2_Pqr}
\begin{align}
P_i(x)&=0_{2\times 2}, \\
q_i(x)&=\frac{o_i-x}{|o_i-x|}, \\
r_i(x)&=2-|x-o_i| 
\end{align}
\end{subequations}
for $i=1,2$. 

It follows that when $[x]_1=0$ and $[x]_2\in ([o_1]_2,[o_2]_2)$, the feasible set $\mathcal{U}(x)$ can be rewritten as
\begin{align}
\mathcal{U}(x) = \{u \in \mathbb{R}^2 : [0;1]^Tu \leq 0, [0;-1]^Tu \leq 0\}. \label{eq_example2_Ux}
\end{align}
Since the vectors $[0;1]$ and $[0;-1]$ are linearly dependent, LICQ does not hold at any feasible point of $\mathcal{U}(x)$ in \eqref{eq_example2_Ux}. 

For comparison, we then consider a safety-critical controller in the form of \eqref{eq_optimization_reshaped} with the primary control input $u_c$ defined in \eqref{eq_simulation1_uc} and the reshaped feasible set $\mathcal{U}_L(x)$ defined in \eqref{eq_reshaped_feasibleset} with $A_L$, $b_L(x)$ and $u_s(x)$ given in \eqref{eq_example2_ALbL}, defined in terms of $q_j(x)$ and $r_j(x)$ for $j=1,2$ given in \eqref{eq_simulation2_Pqr}. Since any two rows of $A_L$ are linearly independent, LICQ holds at any feasible point $u_* \in \mathcal{U}_L(x)$ with the number of active constraints at $u_*$ no larger than $2$. 

\begin{figure}[!ht]
\centering
\includegraphics[width=\linewidth]{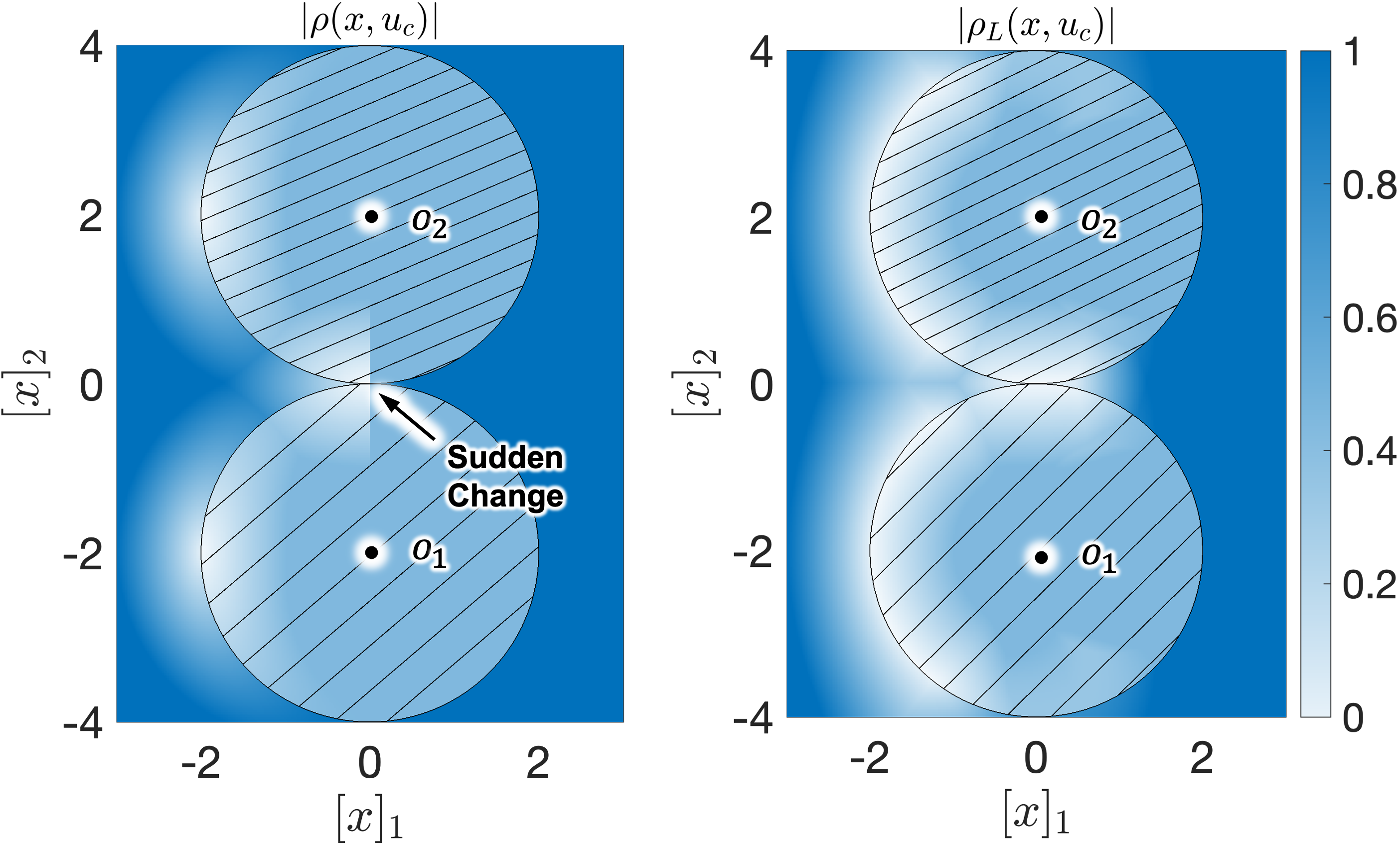}
\caption{Comparison of $|\rho(x,u_c)|$ and $|\rho_L(x,u_c)|$.}
\label{simulation2_compare}
\end{figure}

Denote $\rho(x,u_c)$ and $\rho_L(x,u_c)$ as the maps from $(x,u_c)$ to $u$, generated by solving the QP with the original feasible set $\mathcal{U}(x)$ and the QP with the reshaped feasible set $\mathcal{U}_L(x)$, respectively. Figure \ref{main_ifac03_safety_steps2} illustrates the original feasible set $\mathcal{U}(x)$ and the reshaped feasible set $\mathcal{U}_L(x)$, together with the values of $\rho(x,u_c)$ and $\rho_L(x,u_c)$ for $[x]_1\in\{-2,-1,-0.1,0.1,2\}$ and $[x]_2=0$. It can be observed that the value of $\rho(x,u_c)$ exhibits sudden changes around $[x]_1=0$, while the value of $\rho_L(x,u_c)$ changes continuously. This observation is verified by the difference in the values of $|\rho(x,u_c)|$ and $|\rho_L(x,u_c)|$ near $[x]_1=0$, as shown in Figure \ref{simulation2_compare}.

\section{Conclusion} \label{section_conclusion}

This paper has proposed a feasible-set reshaping technique to ensure that LICQ holds at any feasible point of the reshaped feasible set for which the number of active constraints is no greater than the dimension of the optimization variable. In particular, each quadratic constraint in the original feasible set is represented by half-space constraints. Each half-space constraint is then projected onto the rows of a matrix chosen such that its rows form a positive span of the control input space, and any submatrix with the number of rows no larger than the dimension of the optimization variable is full row rank. The effectiveness of the proposed method has been verified by constructing Lipschitz continuous QP-based controllers based on the reshaped feasible sets. In the future, we will further refine the feasible set reshaping method for more general optimization-based control problems.



\section*{DECLARATION OF GENERATIVE AI AND AI-ASSISTED TECHNOLOGIES IN THE WRITING PROCESS}
No generative AI or AI-assisted technologies were used in the writing process. The authors take full responsibility for the content of the publication.

\bibliography{refs} 

\appendix
\section{Proofs}

\subsection{Proof of Proposition \ref{prop_properties_Ua}} \label{proof_prop_properties_Ua}

We prove the properties one by one. The following proof is carried out for any $x\in\mathcal{X}$.

\textbf{Equivalent representation of $\mathcal{U}(x)$.} Rewrite $\mathcal{U}(x)$ as
\begin{subequations}
\begin{align}
\mathcal{U}(x)&=\bigcap_{i=1,\ldots,n_c}\mathcal{U}_i(x),\\
\mathcal{U}_i(x)&=\{u\in\mathbb{R}^{n_u}:g_i(x,u)\le 0\}.
\end{align}
\end{subequations}
By \eqref{eq_linearized_feasible_set}, it can be verified that $\mathcal{U}(x)=\mathcal{U}_a(x)$ holds if
\begin{align}\label{eq_proog_inclusion}
\mathcal{U}_i(x)\subseteq\bigcap_{\nu\in\mathbb{R}^{n_u}}\mathcal{H}_{i,\nu}(x),
~
\bigcap_{\nu\in\mathbb{R}^{n_u}}\mathcal{H}_{i,\nu}(x)\subseteq\mathcal{U}_i(x).
\end{align}

Consider the first inclusion in \eqref{eq_proog_inclusion}. For any $u_*\in\mathcal{U}_i(x)$, we have $g_i(x,u_*)\le 0$, and since $g_i(x,u)$ is continuously differentiable and convex with respect to $u$, the first-order convexity condition in \cite[Equation (3.2)]{Boyd-Vandenberghe-book-2004} implies that
\begin{align}
0\ge g_i(x,u_*)\ge g_i(x,\nu)+\frac{\partial g_i(x,\nu)}{\partial\nu}(u_*-\nu) \label{eq_proposition_inclusion1}
\end{align}
for all $\nu\in\mathbb{R}^{n_u}$. By the definition of $\mathcal{H}_{i,\nu}(x)$ in \eqref{eq_halfspace}, inequality \eqref{eq_proposition_inclusion1} ensures that $u_*\in\mathcal{H}_{i,\nu}(x)$ holds for all $\nu\in\mathbb{R}^{n_u}$.

Consider the second inclusion in \eqref{eq_proog_inclusion}. For any $u_*\in\bigcap_{\nu\in\mathbb{R}^{n_u}}\mathcal{H}_{i,\nu}(x)$, we have 
\begin{align}
g_i(x,\nu)+\frac{\partial g_i(x,\nu)}{\partial\nu}(u_*-\nu)\le 0
\end{align}
for all $\nu\in\mathbb{R}^{n_u}$. Then, choosing $\nu=u_*$ gives $g_i(x,u_*)\le 0$, so $u_*\in\mathcal{U}_i(x)$.

Therefore, \eqref{eq_proog_inclusion} holds, and hence $\mathcal{U}(x)=\mathcal{U}_a(x)$.

\textbf{Normalized representation of $\mathcal{H}_{i,\nu}(x)$.} Consider $\mathcal{H}_{i,\nu}(x)$ defined in \eqref{eq_halfspace}. By adding $({\partial g_i(x,\nu)}/{\partial\nu})(\nu-u_s(x))$ to both sides of the inequality in \eqref{eq_halfspace}, we obtain
\begin{align}
\frac{\partial g_i(x,\nu)}{\partial\nu}(u-u_s(x)) \le \frac{\partial g_i(x,\nu)}{\partial\nu}(\nu-u_s(x)) - g_i(x,\nu). \label{eq_proof_normalizing}
\end{align}

Under Assumption \ref{assumption_original_feasibleset}, we have $\partial g_i(x,\nu)/\partial \nu$ is nonzero for all $\nu\in\mathbb{R}^{n_u}$ and $x\in\mathcal{X}$. Then, dividing both sides of \eqref{eq_proof_normalizing} by $|{\partial g_i(x,\nu)}/{\partial \nu}|$ yields the inequality in \eqref{eq_halfspace_new}. This is consistent with \eqref{eq_proposition_normalized_representation}. 

\textbf{A feasible point of $\mathcal{U}(x)$.} 
Using the definition of $u_s(x)$ in \eqref{eq_proposition_feasible_point} and the condition $P_i(x)=0$, $q_i^T(x)\neq 0$, we have
\begin{align}
&q_j^T(x) u_s(x) + r_j(x) = |q_j(x)|\Big(\frac{r_j(x)}{|q_j(x)|}+\frac{q_j^T(x)}{|q_j(x)|}u_s(x)\Big) \notag \\
&=|q_j(x)|\Big(\frac{r_j(x)}{|q_j(x)|} + \frac{q_j^T(x)}{|q_j(x)|}\sum_{i=1,\ldots,n_c} \frac{q_i(x)}{|q_i(x)|^2}\min\{-r_i(x), 0\}\Big) \notag \\
&=|q_j(x)|\Big(\min\Big\{\frac{r_j(x)}{|q_j(x)|}, 0\Big\} \notag \\
&~~~+\frac{q_j^T(x)}{|q_j(x)|}\sum_{i=1,\ldots,n_c,i\neq j} \frac{q_i(x)}{|q_i(x)|}\min\Big\{-\frac{r_i(x)}{|q_i(x)|}, 0\Big\}\Big) \notag \\
&\le |q_j(x)|\Big(\min\Big\{\frac{r_j(x)}{|q_j(x)|}, 0\Big\} \notag \\
&~~~-\sum_{i=1,\ldots,n_c,i\neq j} \min\Big\{-\frac{r_i(x)}{|q_i(x)|}, 0\Big\}\Big) \notag \\
&= |q_j(x)|\Big(\min\Big\{\frac{r_j(x)}{|q_j(x)|}, 0\Big\} - \min\Big\{-\frac{r_{i'}(x)}{|q_{i'}(x)|}, 0\Big\}\Big) \notag \\
&\le 0 \label{eq_proposition_feasible}
\end{align}
for any $j=1,\ldots,n_c$, where
\begin{align}
i' = \argmin_{i=1,\ldots,n_c,i\neq j}\left(-\frac{r_i(x)}{|q_i(x)|}\right).
\end{align}
In \eqref{eq_proposition_feasible}, we have used the fact that at most one of $r_1(x),\ldots,r_{n_c}(x)$ is positive for the last equality, and used the fact that
\begin{align}
\min\left\{\frac{r_i(x)}{|q_i(x)|},0\right\} \le \min \left\{-\frac{r_{i'}(x)}{|q_{i'}(x)|}, 0\right\}
\end{align}
for $i,i'=1,\ldots,n_c$, $i\neq i'$ for the last inequality. The two facts are ensured by \eqref{eq_proposition_feasible_point_condition} and $|q_i(x)|\neq 0$ for $i=1,\ldots,n_c$. Inequality \eqref{eq_proposition_feasible} ensures that $u_s(s)\in\mathcal{U}(x)$. This ends the proof of Proposition \ref{prop_properties_Ua}.

\subsection{Proof of Proposition \ref{proposition_halfspace}} \label{appendix_proof_proposition_halfspace}

For notational convenience, we omit the subscripts $i,\nu$ in $\mathcal{H}_{i,\nu}(x)$, $A_{i,\nu}(x)$, and $b_{i,\nu}(x)$ throughout this proof.

For each $x\in\mathcal{X}$ and $b'\in\mathbb{R}^{n_l}_+$, the property $u_s(x)\in\mathcal{U}_{b'}(x)$ can be directly verified by substituting $u_s(x)$ for $u$ into the constraint in \eqref{eq_reshaped_feasibleset}, i.e.,
\begin{align}
A_L(u_s(x)-u_s(x)) = 0 \le b',
\end{align}
which directly implies that $\mathcal{U}_{b'}(x)$ is nonempty.

We now prove $\mathcal{U}_{b'}(x)\subseteq\mathcal{U}_L(x)$ under the condition in \eqref{eq_reshaped_halfspace_condition}. By property (c) of $A_L$, there exists a submatrix $[A_L]_{\mathcal{I}}\in\mathbb{R}^{n_u\times n_u}$\footnote{For any matrix $A\in\mathbb{R}^{m\times n}$ and an index set $\mathcal{I} = \{i_1,\dots,i_k\}\subseteq \{1,\dots,m\}$, the submatrix of $A$ formed by the rows with indices in $\mathcal{I}$ is denoted by $[A]_{\mathcal{I}} = [A_{i_1,:}; \dots; A_{i_k,:}] \in \mathbb{R}^{k\times n}$.}, formed by selecting $n_u$ rows of $A_L$, such that these rows can be positively combined to represent $A(x)$, and such that $[A_L]_{\mathcal{I}}A^T(x)\ge c_A$. This implies that there exists a vector $k\in\mathbb{R}^{n_u}_+$ such that 
\begin{align}
k^T [A_L]_{\mathcal{I}} = A(x). \label{eq_proposition_nonempty_positive}
\end{align}
Then, from the definitions of $\phi(b,A^T)$ and $\mathcal{U}_{b'}(x)$ in \eqref{eq_activation_functions} and \eqref{eq_reshaped_halfspace}, respectively, every $u\in\mathcal{U}_{b'}(x)$ satisfies
\begin{align}
[A_L]_{\mathcal{I}}(u-u_s(x)) 
&\le [b']_{\mathcal{I}} \notag \\
&\le [\phi(b(x),A^T(x))]_{\mathcal{I}} \notag\\
&= [A_L]_{\mathcal{I}}A^T(x)b(x). \label{eq_proposition_nonempty_constraints}
\end{align}
Applying \eqref{eq_proposition_nonempty_positive} and multiplying $k$ on both sides of \eqref{eq_proposition_nonempty_constraints} yield
\begin{align}
A(x)(u-u_s(x)) &= k^T [A_L]_{\mathcal{I}} (u-u_s(x)) \notag \\
&\le k^T [A_L]_{\mathcal{I}} A^T(x)b(x) \notag\\
&= A(x)A^T(x)b(x) = b(x).
\end{align}
This, together with the constraint in \eqref{eq_halfspace_new}, implies that every $u\in\mathcal{U}_{b'}(x)$ also satisfies $u\in\mathcal{U}(x)$. 
This ends the proof of Proposition \ref{proposition_halfspace}.

\subsection{Proof of Theorem \ref{theorem}} \label{appendix_proof_theorem}

We prove the properties in Theorem \ref{theorem} one-by-one.

\textbf{Feasibility and constraint satisfaction.} Under Assumption \ref{assumption_original_feasibleset}, we have $u_s(x)\in\mathcal{U}(x)$ for all $x\in\mathcal{X}$. Based on this, we have $b_{i,\nu}(x)\ge 0$ for all $i=1,\ldots,n_c$, $\nu\in\mathbb{R}^{n_u}$, and $x\in\mathcal{X}$. This, together with the definitions of $b_L$ in \eqref{eq_b_L} and $\phi$ in \eqref{eq_activation_functions}, implies that 
\begin{align}
b_L(x)\ge 0, \qquad \forall x\in\mathcal{X}. \label{eq_theorem_bL_nonnegative}
\end{align}

By the definitions of $\mathcal{U}_L(x)$ in \eqref{eq_reshaped_feasibleset} and $b_L(x)$ in \eqref{eq_b_L}, it follows that 
\begin{align}
b_L(x) \le \phi(b_{i,\nu}(x),A_{i,\nu}^T(x)) \label{eq_theorem_bL_small}
\end{align}
for $x\in\mathcal{X}$ and $\nu\in\mathbb{R}^{n_u}$, where $b_{i,\nu}(x)$ and $A_{i,\nu}^T(x)$ are defined in \eqref{eq_proposition_normalized_representation} and are employed to construct the normalized representation of $\mathcal{H}_{i,\nu}(x)$ stated in Proposition \ref{prop_properties_Ua}. 

Under Assumption \ref{assumption_original_feasibleset}, by applying Proposition \ref{proposition_halfspace}, inequalities \eqref{eq_theorem_bL_nonnegative} and \eqref{eq_theorem_bL_small} ensure that $u_s(x)\in\mathcal{U}_L(x)\subseteq\mathcal{H}_{i,\nu}(x)$ for all $i=1,\ldots,n_c$, $\nu\in\mathbb{R}^{n_u}$ and $x\in\mathcal{X}$. This, together with \eqref{eq_linearized_feasible_set}, implies that $\mathcal{U}_L(x)\subseteq\mathcal{U}(x)$ for all $x\in\mathcal{X}$.

\textbf{LICQ.} This property follows directly from the fact that any $n_u$ rows of $A_L$ are linearly independent, as stated in property (b) of $A_L$.

This ends the proof of Theorem \ref{theorem}.

\subsection{Proof of Theorem \ref{theorem2}} \label{appendix_proof_theorem2}

We prove the properties in Theorem \ref{theorem2} one-by-one.

\textbf{An upper bound.}
We first show that $\rho_L$ is defined for all $x\in\mathcal{X}$ and $u_c\in\mathbb{R}^{n_u}$. In particular, $\mathcal{U}_L(x)$ is convex for every $x \in \mathcal{X}$. Hence, by the projection theorem \cite[Proposition B.11]{Bertsekas-book-1997}, the QP \eqref{eq_optimization_reshaped} admits a unique solution for each $x \in \mathcal{X}$. Thus, $\rho_L$ is uniquely defined for all $x\in\mathcal{X}$ and $u_c\in\mathbb{R}^{n_u}$.

Then, the boundedness property is proved by applying the projection theorem \cite[Proposition B.11]{Bertsekas-book-1997} to the QP \eqref{eq_optimization_reshaped}. In particular, as the solution of the QP \eqref{eq_optimization_reshaped}, $\rho_L(x,u_c)$ satisfies
\begin{subequations} \label{eq_rho_L_upper_proof}
\begin{align}
|\rho_L(x,0)| &\le |u_s(x)| \label{eq_rho_L_upper_proof1} \\
|\rho_L(x,u_c)-\rho_L(x,0)| &\le |u_c| \label{eq_rho_L_upper_proof2} 
\end{align}
\end{subequations}
for all $x\in\mathcal{X}$ and $u_c\in\mathbb{R}^{n_u}$. In particular, \eqref{eq_rho_L_upper_proof1} is guaranteed by $(u_s(x) - \rho_L(x,0))^T(0-\rho_L(x,0)) \le 0$, which follows from \cite[Property (b) of Proposition B.11]{Bertsekas-book-1997} together with $u_s(x)\in\mathcal{U}_L(x)$; \eqref{eq_rho_L_upper_proof2} is guaranteed by \cite[Property (c) of Proposition B.11]{Bertsekas-book-1997}.

From \eqref{eq_rho_L_upper_proof2}, using the triangle inequality \cite[Theorem 3.3]{Apostol-book-1973}, we have
\begin{align}
|\rho_L(x,u_c)| &= |\rho_L(x,0) + \rho_L(x,u_c)-\rho_L(x,0)| \notag \\
&\le |\rho_L(x,0)| + |\rho_L(x,u_c)-\rho_L(x,0)|\notag \\
&\le |u_s(x)| + |u_c|
\end{align}
for all $x\in\mathcal{X}$ and $u_c\in\mathbb{R}^{n_u}$.

\textbf{Lipschitz continuity.}
We employ \cite[Theorem 3.1]{Hager-SIAMControl-1979} to establish the Lipschitz continuity of the QP \eqref{eq_optimization_reshaped}. For $x\in\mathcal{X}$, define a new QP 
\begin{align}
&u' = \argmin_{u} |u-u_c'(x,u_c)|^2 &\text{s.t. } u\in\mathcal{U}'_L(x), \label{eq_new_QP}
\end{align}    
where $u_c'(x,u_c)=u_c-u_s(x)$ and $\mathcal{U}'_L(x)=\{u:A_L u \le b_L(x)\}$. It can be verified that $\mathcal{U}'_L(x)\neq \emptyset$ by using $u_s(x)\in\mathcal{U}_L(x)$. 
Then, 
\begin{align}
\rho_L(x) = u_s(x)+ u', \label{eq_equivalent_form}
\end{align}
where $u'$ is the solution to QP \eqref{eq_new_QP}.

Use the constraints $\breve{A}_L u \le \breve{b}'_L(x)$ in QP \eqref{eq_new_QP} that directly affect the optimal solution; changing them would alter the optimal solution. Note that the number of rows of $\breve{A}_L$ cannot exceed $n_u$, since otherwise the rows of $\breve{A}_L$ would be linearly dependent, contradicting the fact that removing any one of them changes the solution.

For any $x\in\mathcal{X}$ and $u_c\in\mathbb{R}^{n_u}$, $u_c'(x,u_c)$, $A_L$ and $b'_L(x)$ are elements of $\mathbb{R}^{n_u}$, $\{A_L\}$ and $\mathbb{R}^{n_l}_+$, respectively, and all the conditions required by \cite[Theorem 3.1]{Hager-SIAMControl-1979} are checked as follows:
\begin{itemize}
\item $\mathbb{R}^{n_u}$, $\{A_L\}$ and $\mathbb{R}^{n_l}_+$ are convex;
\item for each $x\in\mathcal{X}$, $0\in\mathcal{U}'_L(x)$ and the projection theorem \cite[Proposition B.11]{Bertsekas-book-1997} guarantee the existence and uniqueness of the solution to QP \eqref{eq_equivalent_form}, respectively;
\item property (a) of $A_L$, guarantees the boundedness of $|\breve{A}_L|$;
\item property (b) of $A_L$, i.e., the LICQ property established in Theorem \ref{theorem}, guarantees that there exists $c_{A_L}>0$ such that $\lambda_{\min}^{1/2}(\breve{A}_L\breve{A}_L^T) \ge c_{A_L}$, and thus $|\breve{A}_L^T\xi|\ge c_{A_L}|\xi|,\forall \xi$.
\end{itemize}

Then, the function $\rho_L':\mathcal{X}\times \mathbb{R}^{n_u}\rightarrow\mathbb{R}^{n_u}$ representing the map from $(x,u_c)$ to $u'$ generated by the QP \eqref{eq_new_QP} satisfies
\begin{align}
&|\rho_L'(x_1,u_{c1})-\rho_L'(x_2,u_{c2})|  \notag \\
&\le L(|u_c'(x_1,u_{c1})-u_c'(x_2,u_{c2})|+|b_L(x_1)-b_L(x_2)|), \notag \\
&\le L|u_s(x_1)-u_s(x_2)| \notag \\
&\quad + L|u_{c1}-u_{c2}|+L|b_L(x_1)-b_L(x_2)|
\label{eq_rhoL_rewrite}
\end{align}
for all $u_{c1},u_{c2}\in\mathbb{R}^{n_u}$ and $x_1,x_2\in\mathcal{X}$, where we have used \cite[Theorem 3.1]{Hager-SIAMControl-1979} for the first inequality with  $L>0$, and used the definitions of $u_c'$ and $b_L'$ below \eqref{eq_new_QP} for the second inequality. 

Consequently, by considering the form of $\rho_L$ in \eqref{eq_equivalent_form}, the inequality in \eqref{eq_rhoL_rewrite}, and applying the algebraic properties of Lipschitz functions \cite{Cobzas-Miculescu-Nicolae-book-2019}, one proves \eqref{eq_theorem_Lipschitz}. This ends the proof of Theorem \ref{theorem}.

\end{document}